\documentclass[twocolumn,floatfix,prl,showpacs,superscriptaddress]{revtex4}

\usepackage{float}
\usepackage{graphicx}
\usepackage{bm}
\usepackage{amsmath}
\usepackage{amssymb}
\usepackage{bbold}

\newcommand{\tT}{{\tilde T}}
\newcommand{\bH}{{\bar H}}
\newcommand{\bgamma}{{\bar \gamma}}
\newcommand{\beq}{\begin{eqnarray}}
\newcommand{\eeq}{\end{eqnarray}}
\newcommand{\beqq}{\begin{eqnarray*}}
\newcommand{\eeqq}{\end{eqnarray*}}

\begin{document}

\title{All Majorana Models with Translation Symmetry are Supersymmetric}

\author{Timothy H. Hsieh}
\email{thsieh@kitp.ucsb.edu}
\affiliation{Kavli Institute for Theoretical Physics, University of
California, Santa Barbara, CA 93106, USA}
\author{G\'abor B. Hal\'asz}
\affiliation{Kavli Institute for Theoretical Physics, University of
California, Santa Barbara, CA 93106, USA}
\author{Tarun Grover}
\affiliation{Department of Physics, University of California at San
Diego, La Jolla, CA 92093, USA}
\affiliation{Kavli Institute for
Theoretical Physics, University of California, Santa Barbara, CA
93106, USA}

 \pacs{71.10.Fd, 71.10.Pm, 73.20.-r, 11.30.Pb}

\begin{abstract}
We establish results similar to Kramers and Lieb-Schultz-Mattis theorems but involving only translation symmetry and for Majorana modes.  In particular, we show that all states are at least doubly degenerate in any one and two dimensional array of Majorana modes with translation symmetry, periodic boundary conditions, and an odd number of modes per unit cell.  Moreover, we show that all such systems have an underlying $\mathcal{N}=2$ supersymmetry and explicitly construct the generator of the supersymmetry.
  Furthermore, we establish that there cannot be a unique gapped ground state in such one dimensional systems with anti-periodic boundary conditions.  These general results are fundamentally a consequence of the fact that translations for Majorana modes are represented projectively, which in turn stems from the anomalous nature of a single Majorana mode.  An experimental signature of the degeneracy arising from supersymmetry is a zero-bias peak in tunneling conductance.

\end{abstract}

\maketitle

A Majorana mode is a strange concept.  Formally, it
represents $\sqrt{2}$ degree(s) of freedom because two Majorana
modes constitute a single qubit or spinless fermion.  By
construction, the Majorana mode is its own anti-mode: its creation
and annihilation operators are identical \cite{majorana}.  While the
mathematical existence of Majorana modes arises simply from a change
of basis in the particle-hole space, the physical manifestations of
the above properties are extremely nontrivial \cite{wilczek} and
important for quantum computation purposes \cite{kitaev1, kitaev2,
nayak}. The fact that a Majorana is only a fraction of a physical
electron or qubit suggests the possibility of encoding information
in two widely separated Majoranas, each of which is immune to local
decoherence.  Furthermore, the Hermitian nature of the Majorana mode
forces it to exist at zero energy in the superconducting gap of
physical systems, allowing experimentalists to zero in on finding a
zero bias peak in tunneling, which is necessary if a Majorana mode
is present.

There has been tremendous effort \cite{alicea, beenakker, read,
ivanov, volovik} toward realizing these Majorana modes at the
endpoints of both one dimensional topological superconductors in
nanowires \cite{sarma1, oppen, lee, delft, marcus} and atomic chains
\cite{yazdani}, as well as from two dimensional interfaces between
topological insulators and superconductors \cite{fukane,
williams2012, harlingen}.  The increasingly compelling evidence for
single Majorana modes and the substantial activity in this field
suggest that scaling the system to realize multiple Majoranas along
a line or in a two-dimensional grid may be realized in the near
future.  In such setups involving lattices of emergent Majorana
modes, the low energy effective Hamiltonian involves interactions
between such modes, and such models host abundant and fascinating
phenomenology. For example, two \cite{plaquette} and three
dimensional \cite{haahmajorana} lattices of Majorana zero modes may
provide new architectures for quantum information processing and new
topological phases.

%Even in zero dimensions, randomly interacting Majorana
%fermions provide an exact model of holography \cite{kitaev3,
%sachdev}.

%Moreover, there have been several proposals for realizing supersymmetric critical points from a one dimensional chain of Majorana modes \cite{grover, rahmani}.

%With this motivation from rapidly progressing experiments and theoretical advances, it is worth deducing what are the general properties of such interacting Majorana mode Hamiltonians.  For example, for spin systems with appropriate symmetries such as time-reversal and continuous (e.g. $U(1)$) rotation, there are respectively the Kramers \cite{kramers} and Lieb Schultz Mattis (LSM) \cite{lsm} theorems, which place general constraints on the energy spectrum.

Moreover, there have been several proposals which employ Majorana
modes for realizing supersymmetry, which is a highly appealing
concept from particle physics \cite{susy1, susy2, susy3, susy4},
relating bosonic and fermionic modes to each other.  Though
signatures of it have yet to be observed, there are several
condensed matter systems in which supersymmetry may emerge at long
time and distance scales (``scaling limit''), especially close to a
critical point \cite{balents, sungsik, grover, ponte, yao}.  In
particular, the supersymmetric tricritical Ising model may be
realized at a critical point of Majorana systems \cite{grover,
rahmani}.  Furthermore, time-reversal acts as a supersymmetry on
vortices of topological superconductors \cite{qi}.  However, exact
supersymmetry in lattice models typically requires fine-tuned
Hamiltonians \cite{fendley, feguin, huijse1, huijse2, bauer, yu}.

In this work, we show that {\it all} Majorana systems with
translation symmetry and an odd number of Majorana modes per unit
cell exhibit $\mathcal{N}=2$ supersymmetry and we explicitly
construct its generator and identify its experimental consequences.  For one dimensional systems with periodic
boundary conditions, we establish a Kramers-like theorem
\cite{kramers} but for translation, not time-reversal symmetry: we
show that every energy level is at least doubly degenerate.  With
anti-periodic boundary conditions, we establish a result along the
lines of Lieb-Schultz-Mattis \cite{lsm} and rule out the possibility
of a unique gapped ground state, using recent results for spin
chains \cite{xiechen, mike, cirac, sid}.  For two dimensional systems, we
also establish at least twofold degeneracy for all states and for
all system dimensions.  The essence of all these results is the
fractional nature of the Majorana mode. Each unit cell, with an odd
number of Majoranas, cannot exist intrinsically, and therefore the
symmetry group involving translations and fermion parity is
represented projectively.  We will now illustrate this in detail and conclude by mentioning several experimental venues for our results, in which a striking signature of supersymmetry is a zero-bias peak in tunneling experiments.

\section{1D, periodic boundary conditions}

\begin{figure}
\centering
\includegraphics[height=1.4in]{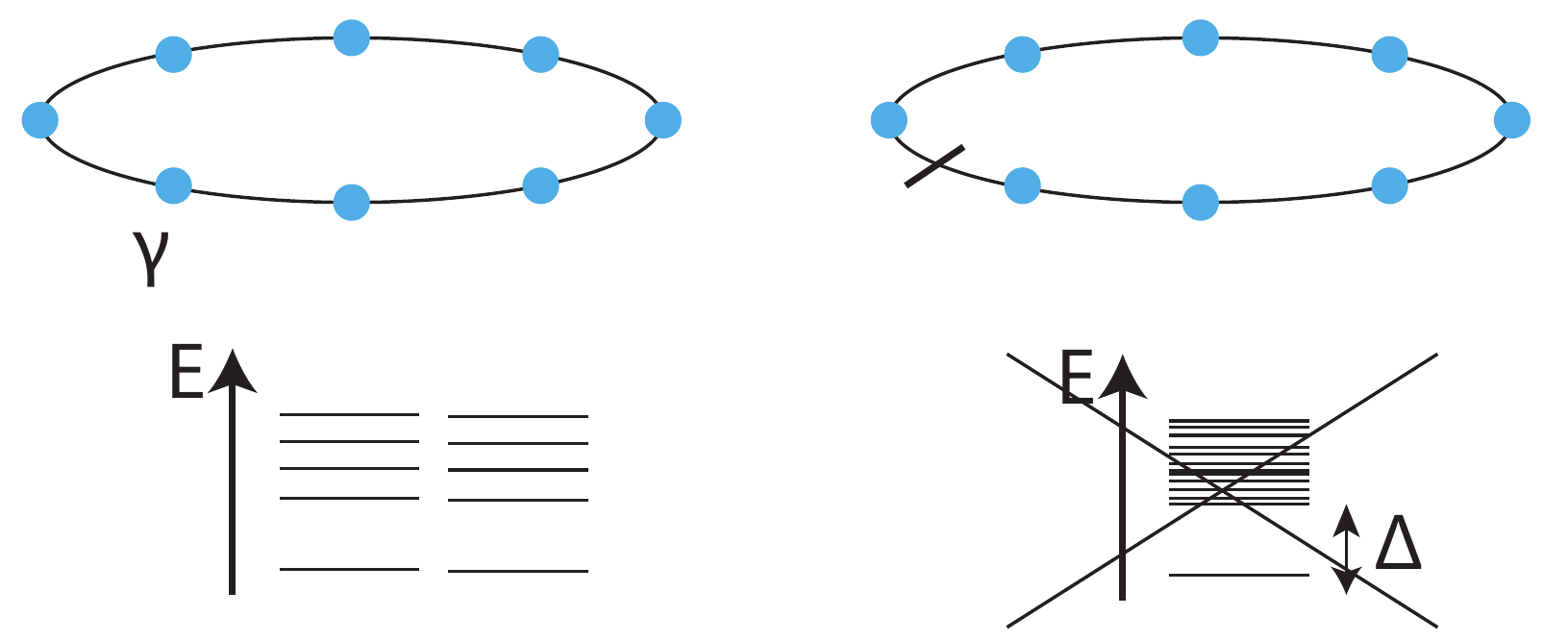}
\caption{(left) Translationally invariant Majorana modes, with
periodic boundary conditions, have at least twofold degeneracy in
the energy spectrum.  The underlying supersymmetry requires that
each energy level contains pairs of fermionic and bosonic
superpartners. (right) The same system, with anti-periodic boundary
conditions (depicted by a slash through a bond), cannot have a
unique gapped ground state in the thermodynamic limit.}
\end{figure}

Consider $N$ (even, to have a well-defined total Hilbert space)
Majorana modes $\{\gamma_i\}_{i=1\cdot\cdot\cdot N}$ localized at
$N$ sites spaced around a ring, and consider general Hamiltonians
invariant under translation by one site and with periodic boundary
conditions: \beq H = \sum_{i=1}^N H_i. \eeq We make no assumptions
on the structure of $H_i$ other than translation invariance and the
conservation of fermion parity. In particular, let $\hat{P}\equiv
i^{N/2}\prod_{i=1}^N \gamma_i$ be the fermion parity operator, for
which we have $[\hat{P},H_i]=[\hat{P},H]=0$.

Let $T$ be translation by one site: \beq \hat{T}\gamma_i \hat{T}^{-1} = \gamma_{i+1 \,
(\mathrm{mod} \, N)}. \eeq By assumption, \beq
\hat{T}H_i \hat{T}^{-1} &=& H_{i+1 \, (\mathrm{mod} \, N)} \\
\implies [\hat{T},H]&=&0. \eeq However, translation and fermion
parity anti-commute: \beq \hat{T}\hat{P}\hat{T}^{-1} =
i^{N/2}\Big(\prod_{i=2}^N \gamma_i\Big) \gamma_1 = -\hat{P} \eeq
because $N$ is even and $\gamma_1$ is anti-commuted through an odd
number of Majoranas to return to the beginning.  We note that the
fact that $P$ and $T$ can anti-commute has been used to classify
fermionic phases protected by translation symmetry \cite{bahri}. It
follows that every eigenstate of $H$ is at least doubly degenerate:
if $H|\psi\rangle = E|\psi\rangle$ and $P|\psi\rangle =
p|\psi\rangle$ ($p=\pm 1$), then $HT|\psi\rangle = TH|\psi\rangle =
ET|\psi\rangle$ and $PT|\psi\rangle =
-TP|\psi\rangle=-pT|\psi\rangle$, and thus $T|\psi\rangle$ is also
an eigenstate with energy $E$ and orthogonal to $|\psi\rangle$.  A similar algebraic structure (but not involving translation) was used in \cite{jlee} to establish spectrum doubling.

\section{Degeneracy as a consequence of $\mathcal{N}=2$ Supersymmetry}

We now show that all translationally invariant Majorana Hamiltonians
in 1D with periodic boundary conditions and an odd number of
Majoranas per unit cell are supersymmetric. The two-fold degeneracy
of the spectra found in the previous section can then be thought of
as a consequence of this underlying supersymmetry.

We first shift all the eigenvalues of the Hamiltonian $H$ by a
constant so that they are all non-negative. Then we define the
following fermionic, non-Hermitian operator $\hat{Q}$:
\begin{equation}
\hat{Q} = \sqrt{\frac{H}{2}} \hat{T} (\hat{1} + \hat{P}),
\label{eq:Q}
\end{equation}
where $\hat{1}$ is the identity operator. Clearly, $\hat{Q}$
commutes with the Hamiltonian $H$: $[H,\hat{Q}]=0$. Most
importantly, due to the relation $\{\hat{T},\hat{P}\} = 0$, one
finds \beq
&& \hat{Q}^2  =  (\hat{Q^{\dagger}})^2 = 0, \label{eq:q_sqr}\\
&& \{\hat{Q},\hat{Q^{\dagger}}\} = 2 H. \label{eq:qqdagger} \eeq
Therefore, $\hat{Q}$ acts as the generator of an $\mathcal{N} = 2$
supersymmetry ($\mathcal{N}$ equals two because $\hat{Q}$ is a
non-Hermitian operator and can be decomposed as $\hat{Q} = \hat{Q}_1
+ i \hat{Q}_2$ where $\hat{Q}_1$ and $\hat{Q}_2$ are Hermitian) \cite{aside}.
Thus, \textit{all} Majorana Hamiltonians in 1D which have an odd
number of Majoranas per unit cell with periodic boundary conditions
furnish an $\mathcal{N} = 2$ supersymmetry.

Supersymmetry naturally explains the results derived in the previous
section on the nature of spectra. All energy eigenvalues are doubly
degenerate, and the corresponding eigenstates can be chosen as
fermion parity eigenstates with opposite parity. Explicitly, given
an eigenstate $|n\rangle_B$ with energy $E_n$ and fermion parity
$+1$, its fermionic partner eigenstate, which has the same energy
eigenvalue $E_n$ but opposite parity, is given by:

\begin{equation}
|n\rangle_F = \frac{\hat{Q}|n\rangle_B}{\sqrt{2E_n}},
\label{eq:susypartner}
\end{equation}
where we have normalized so that ${}_F\langle
n|n\rangle_F = 1$. The opposite fermion parity of $|n\rangle_F$
follows because $\{\hat{P},\hat{Q}\} = 0$.

Due to the factor of $\frac{1}{\sqrt{2E_n}}$ in
Eq.~(\ref{eq:susypartner}), in a general supersymmetric theory, the
existence of supersymmetric partner eigenstates is guaranteed  only
when $E_n \neq 0$. However, in our case, the zero of energy plays no
special role (recall that, generically, we already have to shift all
energy levels by a constant so that $E_n \geq 0$), and therefore,
the supersymmetric partner eigenstates exist for all $n$, including
the ground state. Therefore, the Witten index, which is defined as
the difference between the number of bosonic and fermionic
groundstates, is zero.

\section{1D, anti-periodic boundary conditions}

Local Hamiltonians of the above type but with anti-periodic boundary conditions commute with the twisted translation operator $\tT$, which has the action
\beq
\hat{\tT}\gamma_i \hat{\tT}^{-1} &=& \gamma_{i+1} \quad (i<N), \\
\hat{\tT}\gamma_N \hat{\tT}^{-1} &=& -\gamma_1. \eeq Since
$\hat{\tT}$ commutes with $\hat{P}$, the degeneracy found above is
not required here.

However, we now show that for such Hamiltonians with anti-periodic
boundary conditions, it is not possible for there to be both a
unique ground state and a finite excitation gap in the thermodynamic
limit.  Such constraints, with origins in the Lieb-Schultz-Mattis
theorem, have been recently established \cite{xiechen, mike} for spin chains in which
each unit cell transforms under a projective representation of a
global symmetry (e.g., time reversal).  We will now make contact
with these recent results by doubling the Majorana system and
reinterpreting it as a spin system with additional symmetry from the
doubling construction.

Assume for the sake of contradiction that the Hamiltonian $H$ has a
unique gapped ground state $|\psi_0\rangle$.  Consider the doubled
system $H_D = H + \bH$ where $\bH$ is simply a second copy
of $H$ with Majorana operators represented by $\bgamma$.  Since each
subsystem has a unique gapped ground state, the composite $H_D$ also
has a unique gapped ground state $|\psi_0\rangle \otimes
|\psi_0\rangle$. We now Jordan-Wigner transform $H_D$ into a spin
system: \beq
\gamma_i = \Big(\prod_{j<i} \sigma^z_j\Big) \sigma^x_i, \\
\bgamma_i = \Big(\prod_{j<i} \sigma^z_j\Big) \sigma^y_i. \eeq Care
must be taken because the spin system with fixed boundary conditions
only corresponds to a fixed fermion parity sector of the fermionic
system.  It is straightforward to check that the spin system with
periodic boundary conditions corresponds to the fermion system with
anti-periodic boundary conditions and even fermion parity.  This
sector includes the composite ground state $|\psi_0\rangle \otimes
|\psi_0\rangle$ (whose parity is the square of the parity of
$|\psi_0\rangle$).

%One can see this pretty quickly: consider any Majorana interaction which crosses the boundary between site $N$ and site 1.  If there are an even number of Majoranas on either side of the boundary, then the Jordan Wigner strings cancel in pairs and the boundary twist above also cancels in pairs.  If there are an odd number on either side, the above logic reduces this case to a two Majorana interaction $i\gamma_a \gamma_b$, where $a,b$ are on different sides of the boundary.  Compared to $a,b$ been on the same side, there is a product of all gammas missing between $\gamma_a$ and $\gamma_b$.  Hence, there is a relative sign $-P$, where the minus comes from anticommuting the product with $\gamma_a$.

Through the doubling construction, $H_D$ has a set of discrete
symmetries involving swapping the two chains.  There is a $Z_2$
group generated by $\gamma \leftrightarrow \bgamma$ and a
$Z_4$ group generated by $
\gamma \rightarrow -\bgamma,
\bgamma \rightarrow \gamma$. In the spin language, these
correspond respectively to the symmetries \beq
\sigma^x_i &\leftrightarrow& (-1)^{i+1} \sigma^y_i, \\
\sigma^z_i &\rightarrow& -\sigma^z_i, \eeq and \beq
\sigma^x_i &\rightarrow& -\sigma^y_i, \\
\sigma^y_i &\rightarrow& \sigma^x_i, \\
\sigma^z_i &\rightarrow& \sigma^z_i. \eeq Altogether, we have an
onsite $D_4$ group of rotations which is represented projectively
(by spin $1/2$).

%Alternatively, one can make the $Z_2$ subgroup site-independent at the cost of making the symmetry anti-unitary:
%\beq
%\sigma^x_i &\leftrightarrow& \sigma^y_i \\
%\sigma^z_i &\rightarrow& \sigma^z_i \\
%i &\rightarrow& -i
%\eeq
%In this case, the symmetry group is $Z_4 \rtimes Z_2^T$.

Hence, the arguments in \cite{xiechen, mike, cirac} rule out a
gapped unique ground state of the spin system; in brief, a gapped unique ground state in one dimension is short-range entangled, and this local structure leads to incompatibility between the projective representation of each unit cell and translation symmetry.  By contradiction,
the original fermion chain cannot have a unique ground state with a
gap in the thermodynamic limit.

%(For a quick way to see this, see Fig. 2 and caption)

\section{Two and Higher Dimensions}

In this section, we first consider two-dimensional systems with
translation symmetry in both directions, periodic boundary conditions, and a single Majorana mode
per unit cell.  If the system has one odd length, then degeneracy of
all energy levels follows by bundling all Majoranas along the odd
length direction into a supercell and applying the 1D argument
above.  However, this method does not apply to systems with two even
dimensions, but nevertheless the degeneracy holds.  The fundamental
reason is the fact, established below, that the two translations
$T_X$ and $T_Y$ along the two directions $X$ and $Y$ anticommute
when both dimensions $N_X$ and $N_Y$ are even.  This implies
(in conjunction with $[\hat{T}_X, H] =[\hat{T}_Y, H]= 0$) that all
states have at least twofold degeneracy.

We label the array of Majorana modes $\gamma_{i,j}$ by their row $i$ and column $j$ positions.  Translation $T_X$ has
projective representation given by the product of translations for each row:
\beq
\hat{T}_X &=& \prod_{r=1}^{N_Y} \hat{T}_{X,r}, \\
\hat{T}_{X,r} &=& \gamma_{r,1} \exp \left[ \frac{1}{4}
\sum_{i,j=1}^{N_X} B_{ij} \gamma_{r,i} \gamma_{r,j} \right]. \eeq
See the Supplementary Material for an explanation of why the translation operator has the above form; the essential feature is the Majorana operator $\gamma_{r,1}$ ($B$, an antisymmetric matrix, is not important for our purposes).
Then \beq
\hat{T}_Y \hat{T}_X \hat{T}_Y^{-1} = \prod_{r=1}^{N_Y} \hat{T}_Y \hat{T}_{X,r} \hat{T}_Y^{-1} = \prod_{r=1}^{N_Y} \hat{T}_{X,r+1 \, (\mathrm{mod} \, N_Y)} \\
=-\hat{T}_X \eeq because each $\hat{T}_{X,r}$ involves an odd number
of distinct Majorana operators and there are an odd number ($N_Y-1$)
of anti-commutations required to return to the original ordering of
$\hat{T}_X$.  Thus, $\{ \hat{T}_X, \hat{T}_Y \} = 0$, which ensures all
states have at least twofold degeneracy.  We note that in this case, the degeneracy is not due to supersymmetry\cite{screw}.

The above results for periodic boundary conditions readily generalize to three dimensional systems
with at least one dimension of odd length, but we note that systems with three even length dimensions need not be degenerate.  As a
simple counterexample, consider a $2\times 2\times 2$ array of
Majorana modes with four-Majorana interactions on each face; this
Hamiltonian has a unique ground state.

\section{Applications and Phenomenology}

Such Hamiltonians involving interacting Majorana modes serve as
effective models for either the boundaries or vortex lattices of
topological superconductors.  For example, a stack of topological
superconducting wires hosts an array of Majorana modes localized at
the ends of the wires (see Fig. 2).  The low energy physics of such
systems is thus described by the interactions of the Majorana modes,
for which our work is relevant.

The particular interactions between Majorana modes depend on how the
wires are coupled to each other, and as an example, a natural
effective Hamiltonian for such systems is considered in
\cite{rahmani, milsted}: \beq H=-it\sum_j \gamma_j \gamma_{j+1} +
g\sum_j \gamma_j \gamma_{j+1} \gamma_{j+2} \gamma_{j+3}.
\label{eq:H} \eeq For a particular ratio of $t/g$, the above system
is in the (supersymmetric) tricritical Ising universality class
\cite{rahmani}.  However, our work demonstrates that for {\it all}
values of $t,g$, the above system exhibits $\mathcal{N}=2$
supersymmetry, and as a consequence all energy levels are at least
doubly degenerate.

\begin{figure}
\centering
\includegraphics[height=1.7in]{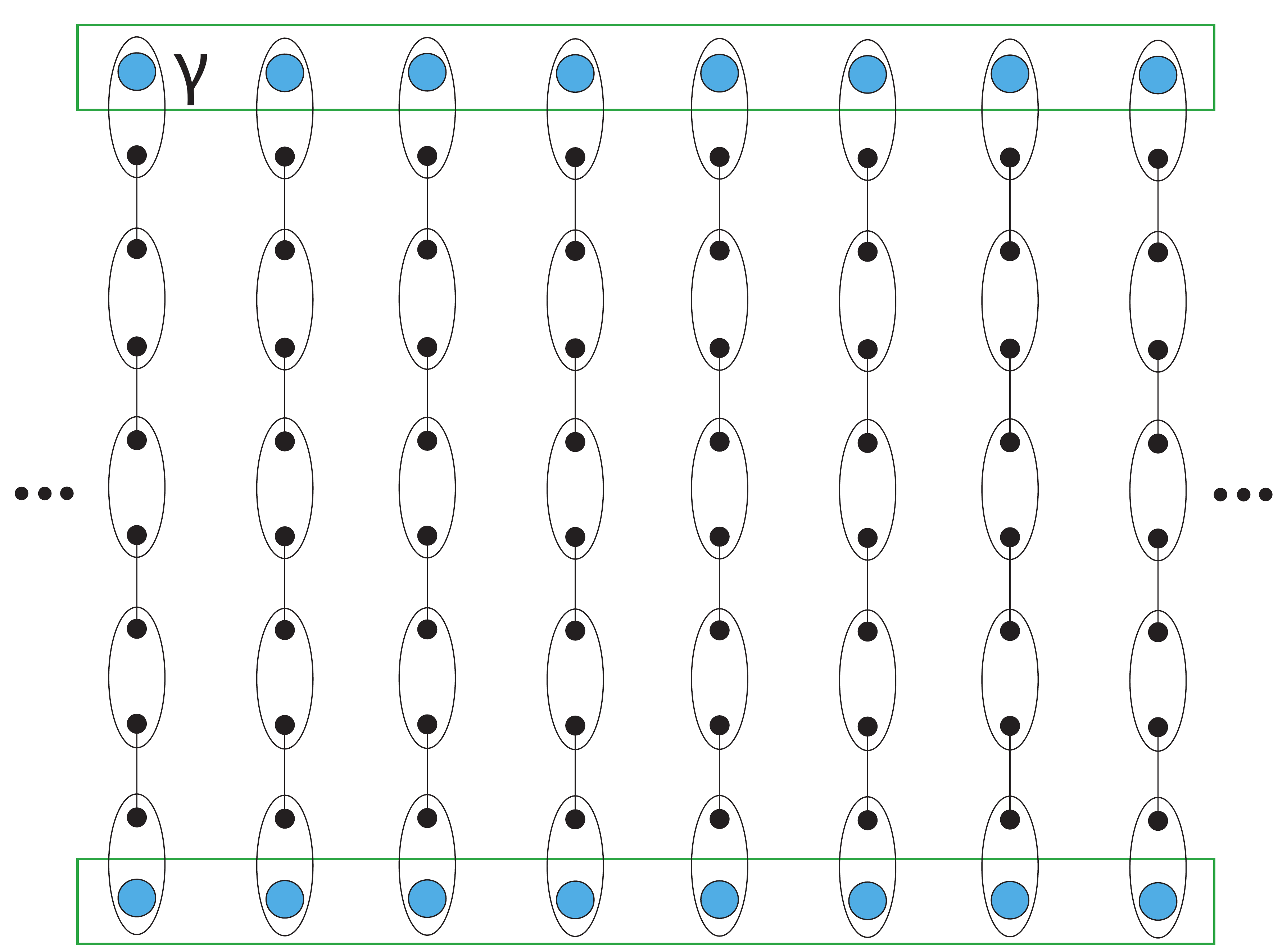}
\caption{Our work applies to effective Hamiltonians (green)
describing the Majorana modes (blue) emerging at the endpoints of
topological superconductors (vertical chains).  Each oval is a
spinless fermion consisting of two Majorana fermions (black dots).}
\end{figure}

Such degeneracy between states of opposite fermion parity has a
distinct signature in tunneling experiments: for a point contact
located near the endpoint of one topological superconducting wire,
there will be a zero bias peak in the tunneling conductance.
However, because the operator $Q$ (which connects a state to its
superpartner) is generically non-local, the zero-bias peak will be
harder to observe as system size grows.  More precisely, for a
point contact to a non-interacting normal lead near the endpoint of wire
$j$, the tunneling conductance is $G \sim e^2 / h$ if the
temperature and the voltage bias are both smaller than $\Lambda^{*}
\propto |{}_F\langle 0 | \gamma_j | 0 \rangle_B|^2$, where $| 0
\rangle_F$ and $| 0 \rangle_B$ are the degenerate ground-state
superpartners \cite{sarma2, lutchyn}.  For the Hamiltonian in
Eq.~(\ref{eq:H}), we find that $|{}_F\langle 0 | \gamma_j | 0
\rangle_B|^2 = 2/N$ for any system size $N$ in the non-interacting
case of $g = 0$. Furthermore, by means of exact diagonalization, we
verify that $|{}_F\langle 0 | \gamma_j | 0 \rangle_B|^2\propto N^{\nu}$ with an exponent $\nu = -1.0 \pm 0.1$ in the range of $8 \leq N
\leq 20$ for all parameter values $|g / t| \leq 1$.  For relatively
small numbers of superconducting wires (which are realistic for
experiments), we therefore expect the zero-bias peak to be
observable.  Note that while the zero-bias peak is expected for a
{\it single} Majorana mode, it is generically not present for an
even number of modes; the translation/supersymmetry is crucial here.

There are other routes toward experimental realization of the many
models of the above type, including Abrikosov vortex lattices on the
surface of topological insulators \cite{chiu} and Josephson-coupled
topological superconductor islands \cite{vijayfu}, in which charging
energy mediates interactions between the emergent Majorana modes.

\section{Summary and Discussion}

We have shown that one and two dimensional systems of Majorana modes
with translation symmetry, periodic boundary conditions, and odd
number of modes per unit cell, have at least twofold degeneracy for
every state in the energy spectrum and that this is a reflection of
an underlying $\mathcal{N}=2$ supersymmetry.  Moreover, we have
shown that such a one dimensional system with anti-periodic boundary
conditions cannot have a unique gapped ground state in the
thermodynamic limit.  Such Majorana systems may be realized at the boundaries or vortex lattices of topological superconductors, and the degeneracy arising from supersymmetry is potentially manifest as a zero-bias peak in tunneling experiments.

Our results motivate the conjecture that for all translationally invariant Majorana systems with an odd number of modes per unit cell, there cannot be a unique gapped ground state in the thermodynamic limit, regardless of dimension or boundary condition.  This leaves the possibilities of gaplessness and symmetry breaking in one dimension, and the additional possibility of topological order in higher dimensions.  Furthermore, while our one-dimensional
anti-periodic boundary conditions doubling analysis conveniently
makes use of recent spin system results, it would be very
enlightening to find a direct proof of the result without having to
double the system.   It is not obvious to us how to apply  the flux
insertion arguments given by Oshikawa \cite{oshikawa} and Hastings
\cite{hastings} in our case since the symmetries are discrete.

Translation symmetry is only one of many crystal symmetries that can
be considered.  Other natural extensions include mirror reflection,
inversion, and perhaps non-symmorphic symmetries as well.  The
effect of these symmetries and their interplay with on-site
symmetries such as time-reversal is an intriguing direction for
future work.  For now, we note as a small extension of our work that
mirror reflection and inversion each anti-commute with fermion
parity if the number of Majoranas that are transformed into other
Majoranas (and not themselves) is $4n+2$ for $n \in \mathbb{Z}$. We
focused on translation symmetry because it enables the simplest
manifestation of supersymmetry.

Finally, we note that the double degeneracy of the full spectrum
discussed in our paper can be thought of as ergodicity breaking that
exists at all temperatures. This is because the degenerate
eigenstates that differ in fermion parity cannot be connected by a
local operator in the thermodynamic limit (note that the operator $\hat{Q}$ in Eq.~(\ref{eq:Q})
is non-local in general). Furthermore, if the system does not
spontaneously break translational symmetry, finite energy density
degenerate eigenstates $|n\rangle_F$ and $|n\rangle_B$ will satisfy
${}_F\langle n | \hat{O} | n \rangle_F = {}_B\langle n | \hat{O} | n
\rangle_B$ and ${}_B\langle n | \hat{O} | n \rangle_F = 0$ for all
local operators $O$, which is reminiscent of topological order
\cite{kitaev2, wen1990}.

\begin{acknowledgements}
{\it Acknowledgements:} We thank Leon Balents for helpful comments.  All authors are supported by a fellowship from the Gordon and Betty Moore Foundation (Grant 4304). TG also acknowledges startup funds from UCSD.
\end{acknowledgements}

\section{Supplementary Material}

Here we review the Majorana representation of rotations in the orthogonal group $O(N)$ and explain the explicit form of the translation operator used in the main text.
Consider $N$ Majoranas labeled by $\gamma_1, \gamma_2, \ldots
\gamma_N$. We will hereafter assume that $N$ is even. A generic rotation in
the space of these $N$ Majoranas is given by $\gamma_i' =
\sum_{j=1}^N R_{ij} \gamma_j$, where $R$ is a real orthogonal matrix
satisfying $R \cdot R^T = 1$. We are interested in projective
representations $\hat{R}$ of $R \in O(N)$ such that $\hat{R}
\gamma_i \hat{R}^T = \gamma_i'$ for all $i$. If $R$ is a proper
rotation such that $\det R = +1$, then it can always be written as
$R = \exp A$, where $A$ is a real antisymmetric matrix. In this
case, the projective representation corresponding to $R$ is
\begin{equation}
\hat{R} = \exp \left[ \frac{1}{4} \sum_{i,j=1}^N A_{ij}
\gamma_i \gamma_j \right].
\end{equation}
If $R$ is an improper rotation such that $\det R = -1$, it can
always be written as a product of a proper rotation $R' = \exp A'$
and a reflection given by
\begin{equation}
S = \left( \begin{array}{ccccc} 1 & 0 & 0 & \cdots & 0 \\ 0 & -1 &
0 & \cdots & 0 \\ 0 & 0 & -1 & \cdots & 0 \\ \vdots & \vdots &
\vdots & \ddots & \vdots \\ 0 & 0 & 0 & \cdots & -1
\end{array} \right).
\end{equation}
Note that $\det S = -1$ because $N$ is even. The projective
representation corresponding to $S$ is simply $\hat{S} = \gamma_1$
because $\gamma_1 \gamma_i \gamma_1 = -\gamma_i$ for any $i \neq 1$
but $\gamma_1^3 = \gamma_1$. The projective representation
corresponding to $R$ is then
\begin{equation}
\hat{R} = \gamma_1 \exp \left[ \frac{1}{4} \sum_{i,j=1}^N
A_{ij}' \gamma_i \gamma_j \right].
\end{equation}

For a one dimensional ring with an even number of Majorana modes, the action of translation by one site on the Majorana modes is:
\begin{equation}
T = \left( \begin{array}{cccccc} 0 & 0 & 0 & \cdots & 0 & 1 \\ 1 & 0
& 0 & \cdots & 0 & 0 \\ 0 & 1 & 0 & \cdots & 0 & 0 \\ \vdots &
\vdots & \vdots & \ddots & \vdots & \vdots \\ 0 & 0 & 0 & \cdots & 0
& 0 \\ 0 & 0 & 0 & \cdots & 1 & 0
\end{array} \right).
\end{equation}
This is an improper rotation when $N$ is even, and therefore the
corresponding projective representation reads
\begin{equation}
\hat{T} = \gamma_1 \exp \left[ \frac{1}{4} \sum_{i=1}^N \sum_{j=1}^N
B_{ij} \gamma_i \gamma_j \right],  \label{eq:T}
\end{equation}
where $B$ is an antisymmetric matrix satisfying $T = S \exp B$. It
can be determined by looking at the eigenvalues and eigenvectors of
$T$, but its precise form is not important for our purposes.


\begin{thebibliography}{10}

\bibitem{majorana}
E. Majorana. Nuovo Cim 14, 171 (1937).

\bibitem{wilczek}
See F. Wilzcek. Nat. Phys. 5, 614 (2009) and references therein.

\bibitem{kitaev1}
A. Kitaev. Phys.-Usp. 44 131 (2001).

\bibitem{kitaev2}
A. Kitaev, Ann. Phys. (Amsterdam) 303, 2 (2003).

\bibitem{nayak}
C. Nayak, S. H. Simon, A. Stern, M. Freedman, S. Das Sarma. Rev. Mod. Phys. 80,
1083 (2008).

\bibitem{read}
N. Read, D. Green. Phys. Rev. B 61, 10267 (2000).

\bibitem{ivanov}
D. A. Ivanov. Phys. Rev. Lett. 86, 268 (2001).

\bibitem{volovik}
M. M. Salomaa, G. E. Volovik. Phys. Rev. B 37, 9298 (1988).

\bibitem{alicea}
J. Alicea, Rep. Prog. Phys., 75 076501 (2012).

\bibitem{beenakker}
C. Beenakker. Annual Review of Condensed Matter Physics 4, 113 (2013).

\bibitem{sarma1}
R. M. Lutchyn, J. D. Sau, S. Das Sarma. Phys. Rev. Lett. 105, 077001 (2010).

\bibitem{oppen}
Y. Oreg, G. Refael, F. von Oppen. Phys. Rev. Lett. 105, 177002 (2010).

\bibitem{lee}
J. Liu, A. C. Potter, K. T. Law, and P. A. Lee. Phys. Rev. Lett. 109, 267002 (2012).

\bibitem{delft}
V. Mourik et al, Science 336, 1003 (2012).

\bibitem{marcus}
A. P. Higginbotham, et.al. Nature Physics 11, 1017 (2015).

\bibitem{yazdani}
S. Nadj-Perge, et.al. Science 346, 6209 (2014).

\bibitem{fukane}
L. Fu and C. L. Kane. PRL 100, 096407 (2008).

\bibitem{williams2012}
J. R. Williams, et al., Phys. Rev. Lett. {\bf{109}},  056803 (2012).

\bibitem{harlingen}
C. Kurter, et.al. Phys. Rev. B 90, 014501 (2014).

\bibitem{plaquette}
S. Vijay, T. Hsieh, and L. Fu. Phys. Rev. X 5, 041038 (2015).

\bibitem{haahmajorana}
S. Vijay, J. Haah, and L. Fu. Phys. Rev. B 92, 235136 (2015).

%\bibitem{kitaev3}
%A. Kitaev. A simple model of quantum holography. KITP Entanglement Program (2015). http://online.kitp.ucsb.edu/online/entangled15/.
%
%\bibitem{sachdev}
%S. Sachdev. Phys. Rev. X 5, 041025 (2015).

\bibitem{susy1}
J. Wess, J. Bagger, Supersymmetry and Supergravity (Princeton Univ.
Press, Princeton, NJ, 1992).

\bibitem{susy2}
J. L. Gervais, B. Sakita, Nucl. Phys. B 34, 632?639 (1971).

\bibitem{susy3}
J. Wess, B. Zumino, Nucl. Phys. B 70, 39 (1974).

\bibitem{susy4}
S. Dimopoulos, H. Georgi. Nucl. Phys. B 193, 150 (1981).

\bibitem{balents}
L. Balents, M.P.A. Fisher and C. Nayak, Int. J. Mod. Phys.
B 12, 1033 (1998).

\bibitem{sungsik}
Sung-Sik Lee, Phys. Rev. B 76, 075103 (2007).

\bibitem{grover}
T. Grover, D. Sheng, and A. Vishwanath. Science 344, 280 (2014)

\bibitem{ponte}
Pedro Ponte, Sung-Sik Lee,  New J. Phys. 16, 013044 (2014).

\bibitem{yao}
S-K. Jian, Y-F. Jiang, and H. Yao. Phys. Rev. Lett. 114, 237001
(2015).

\bibitem{rahmani}
Armin Rahmani, Xiaoyu Zhu, Marcel Franz, Ian Affleck Phys. Rev.
Lett. 115, 166401 (2015).

\bibitem{qi}
X-L. Qi, T. L. Hughes, S. Raghu, and S-C. Zhang. PRL 102, 187001 (2009).

\bibitem{fendley}
P. Fendley, K. Schoutens, and J. de Boer. PRL 90, 120402 (2003).

\bibitem{feguin}
A. Feiguin, et.al. Phys. Rev. Lett. 98, 160409 (2007).

\bibitem{huijse1}
L. Huijse, J. Halverson, P. Fendley, and K. Schoutens, Phys. Rev. Lett. 101, 146406 (2008).

\bibitem{huijse2}
L. Huijse, B. Bauer, and E. Berg. Phys. Rev. Lett. 114, 090404 (2015).

\bibitem{yu}
Y. Yu and K. Yang, Phys. Rev. Lett. 105, 150605 (2010).

\bibitem{bauer}
B. Bauer, L. Huijse, E. Berg, M. Troyer, and K. Schoutens. Phys.
Rev. B 87, 165145 (2013).

\bibitem{kramers}
H. Kramers, Proc. Amsterdam Acad. 33, 959 (1930).

\bibitem{lsm}
E. Lieb, T. Schultz, D. Mattis. Ann Phys 16:407466 (1961).

\bibitem{mike}
H. Watanabe, H. Po, A. Vishwanath, and M. Zaletel. PNAS 112, 47, 14551 (2015).

\bibitem{xiechen}
X. Chen , ZC Gu, XG Wen. Phys Rev B 83:035107 (2011).

\bibitem{cirac}
D. Perez-Garcia, et.al. Phys. Rev. Lett. 100, 167202 (2008).

\bibitem{sid}
See also S. Parameswaran, A. Turner, D. Arovas, and A. Vishwanath. Nat. Phys. 9, 299 (2013).

\bibitem{bahri}
Y. Bahri and A. Vishwanath. Phys. Rev. B 89, 155135 (2014).

\bibitem{jlee}
J. Lee and F. Wilzcek. Phys. Rev. Lett. 111, 226402 (2013).

\bibitem{aside}
One could equally well choose
$\hat{Q}'  = \sqrt{\frac{H}{2}} \hat{T} (\hat{1} - \hat{P})$ as the
supersymmetric generator. $\hat{Q}'$ and $\hat{Q}$ are not
independent because they are unitarily related: $\hat{Q}' =
\hat{T}^{\dagger} \hat{Q} \hat{T}$.

\bibitem{screw}
However, one can consider a ``screw'' boundary condition in which translation $T_Y$ across a boundary is supplemented by a translation in $X$; the array is then effectively a one-dimensional chain in which the head of one column is identified with the tail of the {\it next} column.  For this case, $T_Y$ anticommutes with fermion parity and the system is thus supersymmetric.

\bibitem{milsted}
A. Milsted, et.al. PRB 92, 085138 (2015).

\bibitem{sarma2}
S. Das Sarma, M. Freedman, and C. Nayak. npj Quantum Information 1,
15001 (2015).

\bibitem{lutchyn}
R. M. Lutchyn and J. H. Skrabacz. Phys. Rev. B 88, 024511 (2013).

\bibitem{chiu}
C-K. Chiu, D.I. Pikulin, and M. Franz. PRB 91, 165402 (2015).

\bibitem{vijayfu}
S. Vijay and L. Fu. Physica Scripta, Vol. 2016, T168 (2016).

\bibitem{oshikawa}
M. Oshikawa. Phys. Rev. Lett. 84,1535 (2000).

\bibitem{hastings}
M. Hastings. Phys. Rev. B 69, 104431 (2004).


\bibitem{wen1990}
X.-G. Wen and Q. Niu, Phys. Rev. B 41, 9377 (1990).

\end{thebibliography}
\end{document}